\newif\ifshaphered

\ifdefined\isshaphered
\shapheredtrue
\fi

\documentclass[10pt,journal,compsoc]{IEEEtran}
\ifCLASSOPTIONcompsoc
  \usepackage[nocompress]{cite}
\else
  \usepackage{cite}
\fi

\ifshaphered
\usepackage{longtable}
\newcommand\revision[1]{\textcolor{blue}{#1}}

\newcommand\deleted[1]{\st{#1}}

\else

\newcommand\revision[1]{#1}

\newcommand\deleted[1]{}

\fi

\newcommand*{\textlabel}[2]{%
  \edef\@currentlabel{#1}
  \phantomsection
  #1\label{#2}
}\usepackage{hyperref}

\usepackage{tikz}
\usepackage[misc]{ifsym}
\usepackage{xcolor}
\usepackage{dsfont}
\usepackage{numprint}
\usepackage{subfigure}

\usepackage{xurl}
\usepackage{tabularx}  
\usepackage{booktabs}
\usepackage{amsthm,amsmath,amssymb}
\usepackage[ruled,lined,linesnumbered]{algorithm2e}  
\usepackage{algorithmic}
\usepackage{mathrsfs}
\usepackage{graphicx}
\usepackage{pgfplots}
\usepackage{soul}

\newtheorem{cor}{Corollary}

\theoremstyle{definition}
\newtheorem{defn}{Definition}[section]

\newcommand{\Paragraph}[1]{~\vspace*{0.1\baselineskip}\\\textbf{#1}}
\newcommand{\PParagraph}[1]{~\vspace*{0.1\baselineskip}\textbf{#1}}

\newcommand{\name}{\textsf{FederBoost}}

\newcommand{\Cli}{\mathcal{P}}

\newcommand{\x}{{\mathbf{x}}}
\newcommand{\X}{{\mathbf{X}}}
\newcommand{\y}{{\mathbf{y}}}

\newcommand{\GG}{{\mathsf{G}}}
\newcommand{\HH}{{\mathsf{H}}}

\newcommand{\fig}{\textrm{Figure}}
\newcommand{\rulesep}{\unskip \vrule }

\pgfplotsset{compat=1.18}
\begin{document}

\date{}

\ifshaphered
\input{changelog}
\else
\fi

\title{\bf{\name}: Private Federated Learning for GBDT}


\author{Zhihua Tian, Rui Zhang, Xiaoyang Hou,  
        Lingjuan Lyu,~\IEEEmembership{Member,~IEEE,}\\
        Tianyi Zhang,
        Jian Liu$^{\textrm{\Letter}}$,~\IEEEmembership{Member,~IEEE,}
        and~Kui~Ren,~\IEEEmembership{Fellow,~IEEE}
\thanks{$^{\textrm{\Letter}}$Jian Liu is the corresponding author.}
 \thanks{Zhihua Tian, Zhang Rui, Xiaoyang Hou, Jian Liu and Kui Ren are with Zhejiang University, Hangzhou 310000, China (e-mail: {zhihuat, zhangrui98, jian.liu, kuiren}@zju.edu.cn),  XiaoyangHou42@gmail.com.}
 \thanks{L. Lyu is with Sony AI, Tokyo, 108-0075 Japan. E-mail: lingjuanlvsmile@gmail.com.}
 \thanks{Tianyi Zhang is with Amazon. E-mail: tianyia@amazon.com.}
}


\IEEEtitleabstractindextext{%
\begin{abstract}
Federated Learning (FL) has been an emerging trend in machine learning and artificial intelligence. It allows multiple participants to collaboratively train a better global model and offers a privacy-aware paradigm for model training since it does not require participants to release their original training data.
However, existing FL solutions for vertically partitioned data or decision trees require heavy cryptographic operations. 

In this paper, we propose a framework named $\name$ for private federated learning of gradient boosting decision trees (GBDT). 
It supports running GBDT over both vertically and horizontally partitioned data. 
Vertical $\name$ does  \textit{not} require any cryptographic operation and horizontal $\name$ only requires lightweight secure aggregation.
The key observation is that the whole training process of GBDT relies on the \textit{ordering} of the data instead of the values.

We fully implement $\name$ and evaluate its utility and efficiency through extensive experiments performed on three public datasets.
Our experimental results show that both vertical and horizontal $\name$ achieve the same level of accuracy with  centralized training where all data are collected in a central server;
and they are 4-5 orders of magnitude faster than the state-of-the-art solutions for federated decision tree training;
hence offering practical solutions for industrial applications.

\end{abstract}

}

\maketitle


\section{Introduction}

It is commonly known that big data plays an essential role in machine learning.  
Such big data are typically pooled together from multiple data sources and processed by a central server (i.e., \textit{centralized learning}). 
Now, it becomes troublesome to conduct such activities as 
governments are increasingly concerned with the unlawful use and exploitation of users' personal data.
For example, the European Union has recently enacted General Data Protection Regulation (GDPR), which was designed to give users more control over their data and impose stiff fines on enterprises for non-compliance.
Consequently, service providers become unwilling to take the risk of potential data breaches, and centralized learning becomes undesirable.



\textit{Federated learning} (FL)~\cite{mcmahan17a} addresses this challenge by following the idea of transferring intermediate results of the training algorithm 
instead of the data itself. 
More specifically, it offers a privacy-aware paradigm of model training that does not require data sharing but allows participants to collaboratively train a more accurate global model. 
Since 2017 when it was first proposed by Google~\cite{mcmahan17a},  significant efforts have been put by both researchers and practitioners to improve FL~\cite{stich2018local, konevcny2016federated, DBLP:journals/corr/abs-1811-10792, DBLP:journals/corr/abs-1903-03934, chen2018lag, DBLP:journals/corr/abs-1902-00146,lyu2020privacy}.  
Nevertheless, there are still two problems that remain unsolved in the community:
(1) unable to efficiently handle vertically partitioned data, and (2) unable to efficiently support decision trees.

\begin{figure}[htbp]
\centering
\subfigure[Horizontally partitioned data.]{
\label{fig:horizontalFL}
\includegraphics[width=0.22\textwidth]{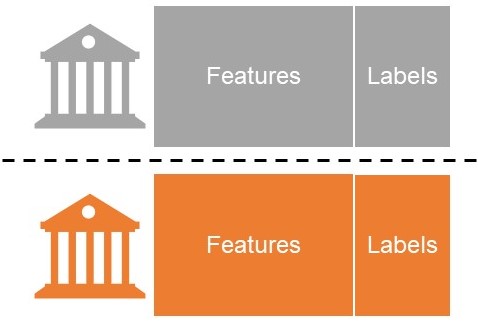}
}
\subfigure[Vertical partitioned data.]{
\label{fig:verticalFL}
\includegraphics[width=0.22\textwidth]{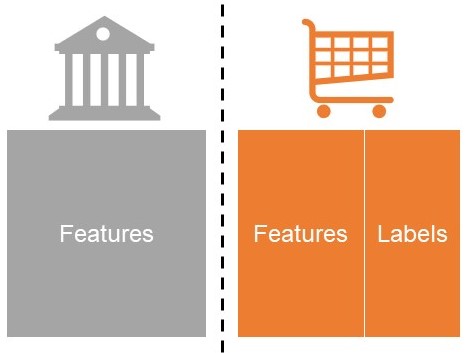}
}
\caption{Data partitions for federated learning.}
\end{figure}

\Paragraph{Horizontal and Vertical FL.}
\revision{\textlabel{Based}{RA:3:1}} on how data is partitioned, \textit{FL} can be roughly classified into two categories: horizontal FL and vertical FL~\cite{yang2019federated}. 
\textit{Horizontal FL}, also known as sample-wise FL,
targets the scenarios where participants' data have the same feature space but differ in samples (cf. \fig~\ref{fig:horizontalFL}).
For example, two regional banks might have the same feature space as they are running the same business; 
whereas the intersection of their samples is likely to be small since they serve different customers in their respective regions. 
\textit{Vertical FL} or feature-wise FL targets the scenarios where participants' data have the same sample space but differ in features (cf. \fig~\ref{fig:verticalFL}). 
For example, consider two participants; one is a bank, and the other is an e-commerce company. 
They can find a large intersection between their respective sample spaces because a customer needs a bank account to use the e-commerce service.
Their feature spaces are certainly different as they are running different businesses:
the bank records users' revenue, expenditure behavior, and credit rating, and the e-commerce company retains users' browsing and purchasing history.
Unlike horizontal FL, which has been extensively studied by the research committee, less attention has been paid to vertical FL. 
Existing vertical FL schemes rely on heavy cryptographic technologies such as homomorphic encryption and secure multiparty computation to combine the feature space of multiple participants~\cite{Hardy2017PrivateFL, yang2019federated, FATE,DBLP:conf/sigmod/FuSYJXT021}.

\Paragraph{Decision trees.}
The FL community focuses on neural networks and pays less attention to other machine learning models, such as decision trees.
Even though neural networks are the most prevailing models in academia, 
they are notorious for lack of interpretability, which hinders their adoption in some real-world scenarios like finance and medical imaging.
In contrast, the decision tree method is regarded as a gold standard for accuracy and interpretability.

A decision tree outputs a sequence of decisions leading to the final prediction, and these intermediate decisions can be verified and challenged separately. 
\deleted{Additionally}\revision{Furthermore}, \textit{gradient boosting decision trees} (GBDT) such as XGBoost~\cite{XGBoost} and LightGBM~\cite{lightGBM} is regarded as a standard recipe for winning ML competitions\footnote{\url{https://github.com/dmlc/xgboost/blob/master/demo/README.md/\#usecases}}, \revision{\textlabel{and}{RA3:1:2} has been widely used in real-world settings for diverse applications, such as aiding explainability challenges in the financial domain~\cite{BOE_ml}, preventing fraudulent activities~\cite{kpmg}, and facilitating business decision-making~\cite{darvish}.} \revision{\textlabel{Additionally}{RA3:1:1}, the Bank of England survey found that tree-based methods remain the most prevalent techniques in financial institutions in England~\cite{BOE_financial_survey}.}
Unfortunately, decision trees have not received enough attention in FL research. 
To the best of our knowledge, most privacy-preserving FL frameworks for decision trees are fully based on cryptographic operations~\cite{secureboost, Benny, DBLP:journals/iacr/AbspoelEV20, DBLP:journals/pvldb/WuCXCO20}, and they are expensive to be deployed in practice.
For example, the state-of-the-art solution~\cite{DBLP:journals/iacr/AbspoelEV20} takes $\sim$28 hours to train a GBDT in LAN from a dataset that consists of  \numprint{8192} samples and \numprint{11} features.

\Paragraph{Our contribution.} In this paper, we propose a novel framework named $\name$ for private federated learning of decision trees. 
It supports running GBDT over \textit{both horizontally and vertically partitioned} data. 

The \textbf{key observation} for designing $\name$ is that the whole training process of GBDT relies on the \textit{ordering} of the samples in terms of their relative magnitudes.
Therefore, in vertical $\name$, it is enough to have the participant holding the labels collect the ordering of samples from other participants; 
then it can run the GBDT training algorithm in exactly the same way as centralized learning. 
We further utilize 
\textit{bucketization} and \textit{differential privacy} (DP) to protect the ordering of samples:
participants partition the sorted samples of a feature into buckets, which only reveals the ordering of the buckets; we also add differentially private noise to each bucket.
Consequently, vertical $\name$ achieves privacy without using any cryptographic operations.

The case for horizontally partitioned data is tricky since the samples and labels are distributed among all participants: no one knows the ordering of samples for a feature.
To conquer this, we propose a novel method for distributed bucket construction
so that participants can construct the same global buckets as vertical $\name$ even though the samples are distributed.
We also use \textit{secure aggregation}~\cite{bonawitz2017practical} to compute the gradients for each bucket, given that no single party holds the labels.
Both the bucket construction method and secure aggregation are lightweight; hence horizontal $\name$ is as efficient as the vertical one.

We summarize our main contribution as follows:
\newenvironment{packeditemize}{
\begin{list}{$\bullet$}{
\setlength{\labelwidth}{8pt}
\setlength{\itemsep}{0pt}
\setlength{\leftmargin}{\labelwidth}
\addtolength{\leftmargin}{\labelsep}
\setlength{\parindent}{0pt}
\setlength{\listparindent}{\parindent}
\setlength{\parsep}{0pt}
\setlength{\topsep}{3pt}}}{\end{list}}

\begin{packeditemize}
    \item We propose $\name$: {\bf a private federated learning framework for GBDT}.
    It supports  {\bf both horizontally and vertically} partitioned data.
    
    \item In vertical $\name$, we define a \textbf{new variant of DP}, which is more friendly for high-dimensional data and saves much privacy budget in a vertical setting.
    
    \item In horizontal $\name$, we propose {\bf a novel method for distributed bucket construction}. 
    
    \item We evaluate the utility of $\name$ on three public datasets. The results show that it achieves {\bf the same level of accuracy with centralized learning}.
    
    \item We provide a {\bf full implementation} of $\name$ and deploy it on a cluster of up to 32 nodes. 
    The benchmark results show that both vertical and horizontal $\name$ are {\bf 4-5 orders of magnitude faster than the state-of-the-art solutions}~\cite{DBLP:journals/iacr/AbspoelEV20, DBLP:journals/pvldb/WuCXCO20} for federated decision tree training.
    
\end{packeditemize}


\section{Preliminaries}

This section provides the necessary background and preliminaries for understanding this paper.

\subsection{Gradient boosting decision tree (GBDT)}
\label{sec:GBDT}

A decision tree is a tree-like model for machine learning predictions. 
It consists of nodes and edges: 
each internal node represents a ``test'' on a feature; 
each edge represents the outcome of the test;
and each leaf node represents the prediction result.
The path from root to leaf represents a prediction rule.

\begin{figure}[htbp]
\centering
\includegraphics[width=0.46\textwidth]{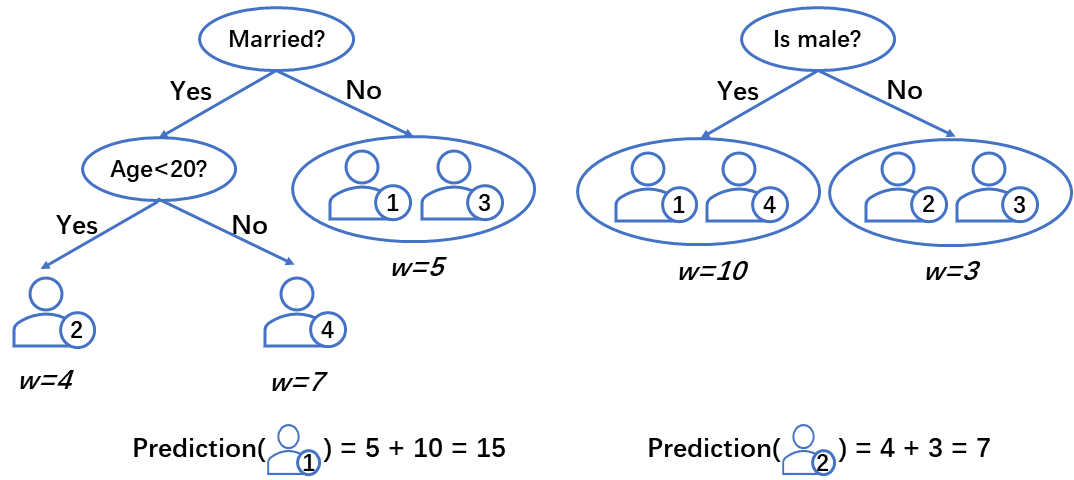}
\caption{An example of GBDT.}
\label{fig:tree}
\end{figure}

Gradient boosting decision tree (GBDT) is a boosting-based machine learning algorithm that ensembles a set of decision trees\footnote{In fact, it is a regression tree; we abuse the notion here.}~\cite{XGBoost}.
Figure~\ref{fig:tree} shows an example of GBDT: in each tree, the input $\x$ is classified to one leaf node that predicts the input with a weight; then it sums the predictions of all trees and gets the final prediction: 
\begin{equation}
\label{equ:prediction}
    \hat{y} = \sum_{t=1}^{T}f_{t}(\x)
\end{equation}
where $T$ denotes the number of trained decision trees and $f_t(\x)$ denotes the prediction result of the $t$th tree.
For classification, it calculates $p = \mathrm{sigmoid}(\hat{y})$ and determines its predicted class based on $p$.

Next, we explain how a GBDT training algorithm works given a dataset $\textbf{X} \in \mathbb{R}^{n\times m}$ that consists of $n$ samples and $m$ features.
It first initializes the prediction result $\hat{y}^0_i$ for each sample with random values.
Then, it trains the $t$th decision tree as follows:
\begin{enumerate}
\item For each sample, calculate the \textit{first-} and \textit{second-order gradient}: 

\begin{equation}
\label{equ:gh}
\begin{split}
g_i=\partial_{\hat{y_i}^{(t-1)}}L(y_i,\hat{y_i}^{(t-1)}), \\
h_i=\partial^2_{\hat{y_i}^{(t-1)}}L(y_i,\hat{y_i}^{(t-1)}) 
\end{split}
\end{equation}

where $\hat{y_i}^{(t-1)}$ is the prediction result aggregated from previous trees, $y_i$ is the real label, and $L(y_i,\hat{y_i}^{(t-1)})$ is the loss function.
The binary cross entropy loss function is typically used as a loss function.


\item Run the following steps for each node of the tree from root to leaf:
\begin{enumerate}
\item For each feature, find the best split of the samples that maximize the following function:
\begin{equation}
\label{equ:split1}
\begin{aligned}
\mathcal{L_\mathit{split}}=&\frac{1}{2} [\frac{(\sum_{i \in I_L} g_i)^2}{\sum_{i \in I_L} h_i +\lambda}\\
+&\frac{(\sum_{i \in I_R} g_i)^2}{\sum_{i \in I_R} h_i +\lambda}+\frac{(\sum_{i \in I} g_i)^2}{\sum_{i \in I} h_i +\lambda}]
\end{aligned}
\end{equation}
where $\lambda$ is a hyper-parameter, $I_L$ denotes the samples divided into the left child node, $I_R$ denotes the samples divided into the right child node, and $I$ denotes all samples in the current node.
The samples in $I_L$, $I_R$, and $I$ are in sorted order of their corresponding feature values.
\item Choose the feature with the maximal $\mathcal{L_\mathit{split}}$ for the current node and split the samples accordingly.
\end{enumerate}
\item The weight of each leaf is computed by the following function:
\begin{equation}
\label{equ:weight}
    w=-\frac{\sum_{i \in I} g_i}{\sum_{i \in I} h_i +\lambda}
\end{equation}
This is the prediction result 
for the samples falling into this leaf.
\end{enumerate}

Most GBDT frameworks accelerate the training process by building a \textit{gradient histogram} for each feature to summarize the gradient statistics; the best split can be found based on the histograms.
More specifically, for each feature, the training algorithm sorts the samples based on their feature values as before. 
Then, it partitions the samples and puts them into $q$ buckets.
For each \textit{bucket}, it calculates $G = \sum_{i \in \mathit{bucket}} g_i$ and $H = \sum_{i \in \mathit{bucket}} h_i$.
The gradient histogram for a feature consists of these $G$s and $H$s of all buckets.
Then, the best split for a feature can be found by maximizing:
\begin{equation}
\label{equ:split2}
\mathcal{L_\mathit{split}}=\frac{1}{2} [\frac{(\sum_{i \in I_L} G_i)^2}{\sum_{i \in I_L} H_i +\lambda}+\frac{(\sum_{i \in I_R} G_i)^2}{\sum_{i \in I_R} H_i +\lambda}+\frac{(\sum_{i \in I} G_i)^2}{\sum_{i \in I} H_i +\lambda}]
\end{equation}
Empirically, 20 buckets are used in popular GBDT frameworks~\cite{DimBoost,vero}.
The split finding algorithm is depicted in Algorithm~\ref{alg:split}.




\renewcommand{\algorithmicrequire}{\textbf{Input}:}
\renewcommand{\algorithmicensure}{\textbf{Output}:}
\renewcommand{\algorithmiccomment}[1]{$\triangleright$ #1}
\begin{algorithm}
\caption{Split Finding}
\label{alg:split}
\small
\KwIn{  $\{G_{1}, ..., G_{q}\}$, $\{H_{1}, ..., H_{q}\}$}
\KwOut {split with max score}
$\GG \gets \sum_{i=1}^qG_{i}$, $\HH \gets \sum_{i=1}^qH_{i}$

$\GG_L \gets 0$, $\GG_R \gets 0$

\For{$i = 1 \to q$}{
$\GG_L \gets \GG_L + G_{i}$,
$\HH_L \gets \HH_L + H_{i}$

$\GG_R \gets \GG - \GG_L$, 
$\HH_R \gets \HH - \HH_L$

$\mathit{score}\gets max(\mathit{score},\frac{G_L^2}{ H_L +\lambda}+\frac{G_R^2}{H_R +\lambda}+\frac{G^2}{ H +\lambda})$
}
\end{algorithm}


\Paragraph{\revision{Prediction.}}
\revision{\textlabel{To}{RA:1:1} generate the prediction of a new sample $\x$ requires running $\x$ on all decision trees and aggregating the output of each tree (cf. equation~\ref{equ:prediction}).
Running a sample on a decision tree includes a sequence of comparison operations. 
Starting at the root node,  the sample is compared to the \textit{threshold value} of the node to determine whether to move it to the left or right child node. 
Then compare the sample with the \textit{threshold value} of the chosen child node to determine whether to move it to the left or right child of the current node. This process repeats until the sample arrives at a leaf node, where the output of the tree is the weight of the leaf.
The threshold value acts as a divider that splits the buckets into two parts. For example, if three buckets $\{\mathit{bucket}_1, \mathit{bucket}_2, \mathit{bucket}_3\}$ was split into two parts:  $\{\mathit{bucket}_1,$ $\mathit{bucket}_2\}$ and $\{\mathit{bucket}_3\}$ on a node, 
the threshold value could be chosen from any value between the maximal value in $\mathit{bucket}_2$ and the minimal value in $\mathit{bucket}_3$ during training.}

\subsection{Federated learning}
\label{sec:fl}

The goal of federated learning (FL) is to enable multiple participants to contribute various training data to train a better model.
It can be roughly classified into two categories~\cite{yang2019federated}: horizontal FL and vertical FL. 

One horizontal FL method~\cite{mcmahan17a} proposed by Google is to distribute the model training process 
of a deep neural network 
across multiple participants by iteratively aggregating the locally trained models into a joint global one. 
There are two types of roles in this protocol: a parameter server and $l$ participants $\Cli$s. 
In the beginning, the parameter server initializes a model  with random values and sends it to all $\Cli$s.
In each iteration, each $\Cli_j$ ($j \in [1,..., l]$) trains the received model  with its local data and sends the parameter server its gradients.
The parameter server aggregates the received gradients and updates the global model.

This elegant paradigm \textit{cannot} be directly applied to vertical FL, where participants have different feature spaces so that
they cannot train models locally.
Furthermore, the FL research committee is focusing on neural networks, and less attention has been paid to decision trees.

\subsection{Secure aggregation}
\label{sec:aggregation}

Bonawitz, et al.~\cite{bonawitz2017practical} propose a \textit{secure aggregation} protocol to protect the local gradients in Google's horizontal FL.
Specifically, they use pairwise additive masking to protect participants' local gradients, 
and have the parameter server aggregate the masked inputs. 
The masks are generated by a pseudorandom generator (PRG) using pairwise shared seeds and will get canceled after aggregation.
The seeds are shared via threshold secret sharing so that dropped-out participants can be handled.
A malicious server can lie about whether a $\Cli_i$ has dropped out, thereby asking all other participants to reveal their shares of $\Cli_i$'s masks.
To solve this issue, they introduce a \textit{double masking} scheme requiring each participant to add another mask to its input and share this mask as well.
The server can request either a share of the pairwise mask (which will get canceled if no one drops) or a share of the new mask;
an honest participant will never reveal both shares for the same participant to the server.
In this paper, we assume the participants are large organizations, and they will not drop out in the middle of the protocol. Thereby, we significantly simplify the secure aggregation protocol.


\subsection{Differential privacy}
\label{sec:dp}


Given a set of input data and an analysis task to perform, the goal of \textit{differential privacy}~\cite{differential} is to permit statistical analysis while  protecting each individual's data.
It aims to ``hide'' some input data from the output: by looking at the statistical results calculated from the input data, one cannot tell whether the input data contains a certain record or not. 

\begin{defn}[$\varepsilon$-Differential Privacy~\cite{differential}]
A randomized algorithm $\mathcal{M}$ with domain $\mathbb{N}^{|\mathcal{X}|}$ is $\varepsilon,$-differentially private if for all $\mathcal{S} \subseteq \mathtt{Range}\left(\mathcal{M}\right)$ and for any neighboring datasets $D$ and $D'$:
\begin{equation*}
    \Pr[\mathcal{M}(D) \in \mathcal{S}] \leq e^\varepsilon\Pr[\mathcal{M}(D')\in \mathcal{S}].
\end{equation*}
\end{defn}

It guarantees that, by examining the outputs $\mathcal{M}(D)$ and $\mathcal{M}(D')$, one cannot reveal the difference between $D$ and $D'$.
Clearly, the closer $\epsilon$ is to $0$, the more indistinguishable $\mathcal{M}(D)$ and $\mathcal{M}(D')$ are, and hence the better the privacy guarantee.
This nice property provides \textit{plausible deniability} to the data owner as the data is processed behind a veil of uncertainty.

\section{Problem Statement}

We consider the setting of $l$ participants $\Cli_1, ..., \Cli_l$, holding datasets $\X_1, ..., \X_l$ respectively, want to jointly train a model.
We consider both vertically (Section~\ref{sec:vertical}) and horizontally (Section~\ref{sec:horizontal}) partitioned data. 
We assume there is a secure channel between any two participants, hence it is private against outsiders.
The participants are incentivized to train a good model (they will \textit{not} drop out in the middle of the protocol), but they want to snoop on others' data.
We do \textit{not} assume any threshold on the number of compromised participants, 
i.e., from a single participant's point of view, all other participants can be compromised. 

\textbf{Poisoning attacks and information leakage from the trained model are not considered in this work.}
We remark that information leakage from the trained model \textit{should be prevented when we consider publishing the model}.
This requires differentially private training~\cite{moments, GBDTAAAI}, which guarantees that one cannot infer any membership about the training data from the trained model. 
However, this line of research is orthogonal to federated learning (which aims to achieve collaborative learning while keeping the training data local), and we leave it as future work to include it in our protocol.


 Given the above setting,  we aim to propose FL schemes with the following design goals:
 \begin{packeditemize}
 \item The {\bf efficiency} should be close to the traditional distributed ML~\cite{XGBoost, vero, DimBoost, lightGBM}, i.e., the number of cryptographic operations should be minimized.
 \item The  {\bf accuracy} should be close to the centralized learning, which is to pool all data into a centralized server.
 \item The  {\bf privacy} should be close to the local training, i.e., each participant trains with its local data only.
To achieve this, all data being transferred should be protected either by cryptographic technology or differential privacy.
\end{packeditemize}

Frequently used notations are summarized in Table~\ref{notationtable}.

\begin{table}[htb]
\small
\centering
\caption{Summary of notations}
\begin{tabular*}{8cm}{r|l}
\toprule
\textbf{Notation} & \textbf{Description} \\
\midrule


$\Cli$       & participant       \\
$l$          & number of participants \\
$\tau$       & number of compromised participants \\
$n$        & number of samples \\
$m$       & number of features\\
$q$       & number of buckets \\
$T$ & number of decision trees\\
$\X$ & dataset \\
$\x^i$ & $i$th feature\\
$\x_j$ & $j$th sample \\
$x^i_j$ & value of $i$th feature $j$th sample \\
$y$ & label \\
$\hat{y}$ & prediction result \\
$g_i$, $h_i$ & first and second order gradient \\
$Q$ & quantile \\
$\epsilon$ & level of differential privacy \\
\bottomrule
\end{tabular*}
\label{notationtable}
\vspace{-6mm}
\end{table}

\section{Vertical {\name}}
\label{sec:vertical}

In vertical FL, $l$ participants $\Cli_1, ..., \Cli_l$, holding feature sets $\X_1, ..., \X_l$ respectively, want to jointly train a model.
Only a single participant (e.g., $\Cli_l$) holds the labels $\y$.
Each feature sets $\X_i$ consists of a set of features: $\X_i = \left[ \x^j, ..., \x^k \right]$ and there are $m$ features in total.
Each $\x^i$ consists all $n$ samples: 

$\x^i =  \left[ x^i_{1}, x^i_{2}, ..., x^i_{n}  \right]^\top$; similarly, $\y = \left[y_{1},y_{2}, ..., y_{n} \right]^\top$
\textbf{We assume that the secure record linkage procedure has been done already}, i.e., all $l$ participants know that their commonly held samples are $\x_{1}...\x_{n}$.
We remark that this procedure can be done privately via \textit{multi-party private set intersection}~\cite{10.1145/3133956.3134065}, which is orthogonal to our paper.

\subsection{Training}

\renewcommand*{\algorithmcfname}{Protocol}
\renewcommand{\algorithmiccomment}[1]{$\triangleright$ #1}
\begin{algorithm}[tb]
\caption{Vertical $\name$}
\label{alg:vertical}
\small
\KwIn {each $\Cli_i$ inputs feature $\x^i = \{x^i_{1}, ..., x^i_{n}\}$ \CommentSty{for simplicity, we assume each $\Cli_i$ holds a single feature, 
hence $m=l$}}
\KwIn {$\Cli_l$ inputs labels $\y = \{y_1, ..., y_n\}$}
\KwOut {$T$ decision trees}

\For(\hfill\CommentSty{for each $\Cli$}){$i = 1 \to l$}{
$\Cli_i$ sorts $\x^i$ and partitions it into $q$ buckets

\For(\hfill \CommentSty{for each sample}){$j = 1 \to n$}{
$\Cli_i$ moves $x^i_{j}$ to another bucket with probability $\frac{q-1}{e^{\varepsilon}+q-1}$
\hfill\CommentSty{add DP noise}
}
$\Cli_i$ sends sample {\bf ID}s in each bucket to $\Cli_l$
}

$\Cli_l$ initializes $\{\hat{y}_1, ...,\hat{y}_n\}$ with random values 

\For(\hfill \CommentSty{for each tree}){$t = 1 \to T$}{
\For(\hfill \CommentSty{for each sample}){$i = 1 \to n$}
{
    $\Cli_l$: $g_i\gets \partial_{\hat{y_i}}L(y_i,\hat{y_i})$, $h_i\gets \partial^2_{\hat{y_i}}L(y_i,\hat{y_i})$
}

\For{{\bf each} node in the current tree}{

\For(\hfill \CommentSty{for each feature}){$i = 1 \to m$}{
\For(\hfill \CommentSty{for each bucket}){$j = 1 \to q$}{

$\Cli_l$ updates $\mathit{bucket}^i_{j}$ \hfill\CommentSty{not for root}

$\Cli_l$: $G^i_{j} \gets  \sum_{k \in \mathit{bucket}^i_{j}}g^i_{j,k}$ $H^i_{j} \gets  \sum_{k \in \mathit{bucket}^i_{j}}h^i_{j,k}$
}
$\mathit{score}_i$, $\mathit{split}_i$ $\gets\Cli_l$ runs {\bf Algorithm~\ref{alg:split}} 
with inputs $\{G^i_{1}, ..., G^i_{q}\}$, $\{H^i_{1}, ..., H^i_{q}\}$
}
for the maximal $\mathit{score}_j$,
$\Cli_l$ sends $\mathit{split}_j$ to $\Cli_j$ 
$\Cli_l$ splits the buckets based on $\mathit{split}_j$
}
$\Cli_l$ updates $\{\hat{y}_1, ..., \hat{y}_n\}$ based on the weights (cf. equation~\ref{equ:weight}) of the leaf nodes. 
}
\end{algorithm}



Vertical $\name$ is based on the observation that the whole training process of GBDT does \textit{not} involves feature values (cf. Section~\ref{sec:GBDT}). Recall that the crucial step for building a decision tree is to find the best split of samples for a feature, which only requires the knowledge of the first- and second- order gradients $g_i$s, $h_i$s, as well as the order of samples (Equation~\ref{equ:split1}).
Furthermore, $g_i$s and $h_i$s are calculated based on the labels and the prediction results of previous trees (equation~\ref{equ:gh}).
Therefore, to train a GBDT model, only the labels (held by $\Cli_l$), the prediction results of the previous tree, and the order of samples are required.

To this end, we let each participant sort its feature samples and tell $\Cli_l$ the order. With this information, $\Cli_l$ can complete the whole training process by itself.
In this way, participants only need to transfer sample orders instead of values, which reveal much less information. 
Another advantage of this method is that the sorting information only needs to be transferred once, and $\Cli_l$ can use it to train the model without further communication.

However, information leakage from the sample orders is still significant.
Take the feature ``salary'' as an example. $ \Cli_l$ can get such information: ``Alice's salary $\leq$ Bob's salary $\leq$ Charly's salary''.
If $\Cli_l$ knows Alice's salary and Charly's salary, it can infer Bob's salary (or at least the range).
We combine two methods to prevent such information leakage: putting samples into buckets and adding differentially private noise (cf. Section~\ref{sec:dp}). In more detail, for each feature, $\Cli_i$ sorts the samples based on their feature values, partitions the samples, and puts them into $q$ buckets.
In this way, $\Cli_l$ only knows the order of the buckets but learns nothing about the order of the samples inside a bucket.
\revision{\textlabel{We note}{RA:2:1} that this partitioning process is uniform for both continuous and categorical data, preventing an adversary from deciphering the feature distribution through an analysis of the number of samples in each bucket.}
To further protect the order of two samples in different buckets, we add differentially private noise to each bucket. 
That is, for a sample that was originally assigned to the $i$th bucket: 
\begin{itemize}
    \item with probability $p= \frac{e^{\varepsilon}}{e^{\varepsilon}+q-1}$, it stays in the $i$th bucket;
    \item with probability $p'=\frac{1}{e^{\varepsilon}+q-1}$, it moves to the $j$th bucket that is picked uniformly at random. 
\end{itemize}
This mechanism is similar to \textit{random response}~\cite{LDP}, which achieves $\epsilon$-LDP.
Let $\mathsf{bucketize}(x)$ denote the bucketization mechanism mentioned above and its output is the bucket ID.
Then, for any two samples $x_1$, $x_2$ and a bucket $B$, we have $\forall k\in \{1,..., q\}$:
\begin{equation*}
\begin{aligned}
    \frac{Pr\left [\mathsf{bucketize}(x_1)= k\right ]}
    {Pr\left [\mathsf{bucketize}(x_2)=k\right ]} \leq&
    \frac{p}{p'} =& \frac{e^{\varepsilon}/(e^{\varepsilon}+q-1)}{1/(e^{\varepsilon}+q-1)}\\
    =& e^{\varepsilon},
\end{aligned}
\end{equation*}
where $Pr\left [\mathsf{bucketize}(x)=k\right ]$ denotes the probability that a sample $x$ is placed in bucket $B_k$. 
We present the security analysis of the mechanism in Section~\ref{sec:valueDP}. 

Our experimental results (cf. Section~\ref{sec:utility}) show that when $\epsilon=4$ and $q=16$, the accuracy achieved by vertical $\name$ is very close to that without DP.
On the other hand, with this configuration, each sample has a probability of approximately 22\% being placed in the wrong bucket. 
The whole training process for vertical $\name$ is depicted in Protocol~\ref{alg:vertical}, and we separately detail the security analysis from the perspective of $\Cli_i$ and $\Cli_l$ in Section~\ref{sec:vertical_analysis}.


\Paragraph{\revision{Prediction.}}
\revision{\textlabel{Recall}{RA:1:2} that the active participant $\Cli_l$ only knows the sample IDs in each bucket; it knows nothing about their values.
However, each $\Cli_i$ holding the values is able to find the \textit{threshold value} after knowing how to split its buckets (line 21 of Protocol~\ref{alg:vertical}).
As a result, the threshold values are distributed among all participants, and they need to run the prediction phase in a distributed way. }
\revision{In more detail, we assume each participant knows some features of $\x$ that correspond to the features it holds during prediction. 
Starting from the root, $\Cli_l$ contacts the participant $\Cli_i$ who holds the corresponding feature;
$\Cli_i$ compares the feature value of $\x$ with the threshold value, and tells $\Cli_l$ the result, based on which $\Cli_l$ decides whether to move left or right. By repeating the process, $\Cli_l$ could get the output of all trees and aggregate the results to make a final prediction.}

\subsection{Element-Level Local DP}
\label{sec:valueDP}

Differential privacy (DP), as well as local differential privacy, provides strong protection against attackers and guarantees that one cannot tell whether a particular sample is in the database or not. However, this kind of strong protection is not necessary for vertical FL, as participants need to know their commonly held samples to run the protocol. In other words, protecting the privacy of data generated by one individual instead of protecting whether one generates private data or not is much more significant. For instance, an online shopping user shopping on the internet many times. In such cases, it may be satisfying from a privacy perspective not to protect whether a user participant in the database but to protect no one knowing any particular \textit{thing} the user has bought. To achieve this goal, more nuanced trade-offs can arise if we wish to prevent an attacker from knowing, for example, whether a user has ever bought a dress.

Here, we focus on local differential privacy. For any different inputs $\x, \x' \in \mathcal{X}$ in the definition of local DP, the word "different" implies that the Hamming distance between $\x,\x'$ is $1$, i.e., $d_{Hamming}(\x,\x'):=\mathds{1}\{\x\neq \x'\}$. This definition makes learning challenging in some scenarios where individual users contribute multiple data items rather than a single item. Thus, a more fine-grained distance notion is needed to keep utility while providing sufficient privacy.


We consider the scenario where data are processed by each individual and propose element-Level Local DP.
We denote the distance between two users' local data $x=[x_1,...,x_k]^\mathrm{T}$ and $x'=[x'_1,...,x'_k]^\mathrm{T} $ is the number of different elements of them, that is,
$$
\begin{aligned}
d_{element}(\x,\x') = & d([x_1,...,x_k]^\mathrm{T}, [x'_1,...,x'_k]^\mathrm{T})\\
=:& \sum_{k=1}^K \mathds{1}\{x_i \neq x'_i\}.
\end{aligned}
$$
Then two users' data $\x,\x'$ are \textit{element-different} if the distance between them $d_{element}(\x,\x')\leq 1$. The definition of local element-level differential privacy is now immediate as follows.
\begin{defn}
[$\varepsilon$-Local Element-Level DP]
An algorithm $\mathcal{M}$ satisfies $\varepsilon$-local element-level differential privacy if for all $y\in \mathtt{Range}(\mathcal{M})$ and for any inputs $\x, \x'$ satisfying $d_{element}(\x,\x')\leq 1$:
$$
\operatorname{Pr}[\mathcal{M}(\x) =y] \leq e^{\varepsilon} \operatorname{Pr}\left[\mathcal{M}\left(\x^{\prime}\right)=y\right].
$$
\end{defn}

Element-level local differential privacy guarantees that the release of a user's data perturbed by a mechanism does not leak any particular "element" the user has. Next, we prove that our bucketize mechanism satisfies element-level local differential privacy, hence providing a sufficient privacy guarantee in our vertical \name.

\begin{cor}
Our bucketization mechanism satisfies $\varepsilon$-element-level local differential privacy.
\end{cor}
\begin{proof}
For any inputs $\x,\x'\in\mathcal{X}$ satisfying $d_{element}(x,x')\leq 1$ and for any $y \in \mathtt{Range}(bucketize)$, we have
\begin{equation*}
\begin{aligned}
    \frac{\Pr[bucketize(\x) = y]}{\Pr[bucketize(\x')=y]}=\prod_i \frac{\Pr[bucketize(x_i) = k_i]}{\Pr[bucketize(x'_i)=y_i]}
\end{aligned}
\end{equation*}
As $\x, \x'$ satisfy $d_{element}(\x,\x')\leq1$, which implies that $x$ and $x'$ differ in only one element (e.g., $x_j\neq x'_j$), we get
\begin{equation*}
\begin{aligned}
    \prod_i \frac{\Pr[bucketize(x_i)=y_i]}{\Pr[bucketize(x'_i)=y_i]}=&\frac{\Pr[bucketize(x_j)=y_j]}{\Pr[bucketize(x'_j)=y_j]}\\
    \leq&e^\varepsilon
\end{aligned}
\end{equation*}
\end{proof}

\subsection{Security analysis}
\label{sec:vertical_analysis}
\PParagraph{$\Cli_i$'s security.} The active participant $\Cli_l$ \textit{only} learns the order of buckets for each feature behind a veil of uncertainty:
each sample has a probability of $p=\frac{q-1}{e^{\epsilon}+q-1}$ to be placed in a wrong bucket.
When $\epsilon={4}$ and $q=16$, $p$ is approximately ${22}\%$.



\Paragraph{$\Cli_l$'s security.}
The only information a passive party $\Cli_i$ learns is the split information sent by $\Cli_l$ (line 21 of protocol 2),
which indicates how the buckets are split into left and right, leading to the maximal $L_{split}$ in equation 2.5. 
Note that the split information of each node made up the final trained decision tree, and information leakage from it is not considered in this work.

Suppose $\Cli_l$ also holds a feature $\x^l$.
If $\Cli_l$'s feature $\x^l$ was selected by a tree node, 
the holders of child-node learn that the samples assigned to left is smaller than the samples assigned to right.
Nevertheless, such information is also protected by differential privacy.



\section{Horizontal $\name$}
\label{sec:horizontal}

In horizontal FL, the dataset is partitioned horizontally:
$l$ participants $\Cli_1, ..., \Cli_l$ hold sample sets $\X_1, ..., \X_l$ respectively. 
Each sample set $\X_i$ consists of a set of samples: 
$\X_i = \left[\x_j, ...,\x_{k} \right]^\top$, and each sample $\x_i$ has all features \textit{and the label}: $\x_i = \left [ x^1_{i}, x^2_{i}, ..., x^m_{i}, y_i \right ]$. 

In this setting, it is natural to have participants train their models locally and aggregate the locally trained models into a joint global model (e.g., Google's FL framework~\cite{mcmahan17a} we mentioned in Section~\ref{sec:fl}). 
This idea applies to decision trees as well: each participant locally trains a decision tree and 
all decision trees are integrated into a random forest via bagging~\cite{RF}.
However, to train a random forest, each participant is required to hold at least 63.2\% of the total samples~\cite{Breiman1996}, 
which contradicts the setting of FL.

We follow the idea of vertical $\name$ to have participants 
jointly run the GBDT training algorithm.
However, there are two challenges we need to conquer in the setting of horizontal FL. 
Firstly, the samples are distributed among $l$ participants, thereby no single participant knows the order of samples for any feature.
To address this, we propose a novel method for distributed bucket construction. 
Secondly, each participant only holds part of the labels, hence the histogram for each bucket is difficult to compute without information leakage.
We have participants calculate $g$s and $h$s locally and aggregate them using secure aggregation (cf. Section~\ref{sec:aggregation}) to build the histograms. 
We provide all details in the rest of this section.

\renewcommand*{\algorithmcfname}{Protocol}
\begin{algorithm}[tb]
\caption{Quantile lookup (for a single feature)}
\label{alg:quantile}
\small
\KwIn{sample values $\{x_1, ..., x_n\}$ of a feature, distributed across $l$ participants} 
\KwOut {quantiles $\{Q_1, ..., Q_{q-1}\}$}
\BlankLine
\For(\hfill \CommentSty{for each quantile}){$j=1 \to q-1$}{
    $\Cli_l$ sets $Q_{min}$ and $Q_{max}$ as the smallest and largest possible values of this feature
    
    $\Cli_l$: $n'\gets 0$ 
    
    \While{$|n' - \frac{n}{q}|> 0 $ }{
        $\Cli_l$: $Q_j \gets \left(Q_{min}+Q_{max}\right)/ 2$, and sends $Q_j$ to other $\Cli$s
        
        \For(\hfill\CommentSty{for each $\Cli$}){$i = 1 \to l$}{
            $\Cli_i$: finds $n'_i$, the total number of local $x$s that are smaller than $Q_j$
        }
        all $\Cli$s run secure aggregation: $n' \gets \sum_{i=1}^{l}n'_i$
        
        \uIf{$n'>\frac{n}{q}$}{
            $\Cli_l$: $Q_{max} \gets Q_j$
        }\ElseIf{$n'<\frac{n}{q}$}{
            $\Cli_l$: $Q_{min} \gets Q_j$}
        }
    \For(\hfill \CommentSty{for each $\Cli$}){$i = 1 \to l$}{
        $\Cli_i$: remove the local $x$s that are smaller than $Q_j$
    }
}
\end{algorithm}

\renewcommand*{\algorithmcfname}{Protocol}
\begin{algorithm}[htb]
\caption{Horizontal $\name$}
\label{alg:horizontal}
\small
\KwIn{each $\Cli_i$ inputs $n_i$ samples and each sample has all $m$ features and a label $y$
}
\KwOut{$T$ decision trees}

\For(\hfill \CommentSty{for each feature}){$i = 1\to m$}{
    $\{Q^i_{1},...,Q^i_{q-1}\}\gets$ all participants run {\bf Protocol~\ref{alg:quantile}}
}

\For(\hfill \CommentSty{for each $\Cli$}){$i = 1\to l$}{
    $\Cli_i$ initializes $\{\hat{y}\}_{n_i}$ with random values
}

\For(\hfill \CommentSty{for each tree}){$t= 1 \to T$}{
    \For(\hfill \CommentSty{for each $\Cli$}){$k = 1 \to l$}
    {
        $\Cli_k$ computes $g_i\gets \partial_{\hat{y_i}}L(y_i,\hat{y_i})$, $h_i\gets \partial^2_{\hat{y_i}}L(y_i,\hat{y_i})$ for all of its samples
    }

    \For{{\bf each} node in the current tree}{
    
        \For(\hfill \CommentSty{for each feature}){$i = 1 \to m$}{
            
            \For(\hfill \CommentSty{for each $\Cli$}){$k = 1\to l$}{
                $\Cli_k$ builds $q$ buckets via $\{Q^i_{1},...,Q^i_{q-1}\}$
            }
            
            \For(\hfill \CommentSty{for each bucket}){$j = 1 \to q$}{
                 \For(\hfill \CommentSty{for each $\Cli$}){$k = 1 \to l$}{
                    $\Cli_k$ computes $G^i_{j,k}$, $H^i_{j,k}$ for $\mathit{bucket}^i_{j,k}$
                 }
                all $\Cli$s run {\bf secure aggregation}: $G^i_{j} \gets \sum_{k=1}^l G^i_{j,k}$, $H^i_{j} \gets \sum_{k=1}^l H^i_{j,k}$
            }
                
            $\mathit{score}_i$, $\mathit{split}_i$ $\gets\Cli_l$ run {\bf Algorithm~\ref{alg:split}} for split finding, with inputs $\{G^i_{1}, ..., G^i_{q}\}$, $\{H^i_{1}, ..., H^i_{q}\}$
        }
        for the maximal $\mathit{score}_{j}$, $\Cli_l$ sends  $\mathit{split}_j$ to other $\Cli$s
        
        all $\Cli$s split the buckets based on $\mathit{split}_j$
    }
    
    update $\{\hat{y}_1, ..., \hat{y}_n\}$ based on the weights (cf. equation~\ref{equ:weight}) of the leaf nodes. 
    
}
\end{algorithm}
\vspace{-4mm}

\renewcommand*{\algorithmcfname}{Protocol}
\begin{algorithm}[htb]
\caption{Secure aggregation (for $G$s)}
\label{alg:aggregation}
\small
\KwIn{each $\Cli_i$ inputs $G_i$}
\KwOut {$G =\sum^l_{i=1}G_i$}
\BlankLine

\CommentSty{\bf Setup:} \hfill\CommentSty{once and for all rounds} \\
$\Cli_l$ initializes a large prime $p$, a group generator $g$, and a small modular $N$; ~sends them to other $\Cli$s

\For(\hfill \CommentSty{for each $\Cli$}){$i=1 \to l-1$}{
    $\Cli_i$: ~~$s_i\overset{\$}{\leftarrow}\mathbb{Z}_{p}$;~~
    $S_i\gets g^{s_i}\mod p$; ~~sends $S_i$ to $\Cli_l$
    
    \For(\hfill \CommentSty{for each $\Cli_j \neq \Cli_i$}) {$j = 1 \to l-1~\mathit{and}~j \neq i$} {
        $\Cli_l$ sends $S_i$ to $\Cli_j$
    }
}

\BlankLine
\CommentSty{\bf Aggregation ($k$th round)}

\For(\hfill \CommentSty{for each $\Cli$}){$i=1 \to l-1$}{
    
    \For(\hfill \CommentSty{for each $\Cli_j \neq \Cli_i$}) {$j = 1 \to l-1~\mathit{and}~j \neq i$} {
        $\Cli_i$: $S_{i,j} \gets S^{s_i}_j \mod p$;
        ~~~~~~~~~~~~~~~~~~~~~~$r_{i,j}\gets \alpha\cdot \mathsf{PRG}(k || S_{i,j})\mod N$ $(\mathit{if}~i < j,~\alpha = 1,~\mathit{else}~\alpha = -1)$\hfill \CommentSty{can be preprocessed}
    }
    $\Cli_i$: $\tilde{G}_i \gets (G_i + \sum_{j\neq i}r_{i,j}) \mod N$; ~~sends $\tilde{G}_i$ to $\Cli_l$
}

$\Cli_l$: $G \gets (\sum^{l-1}_{i=1}\tilde{G}_i + G_l) \mod N$


\end{algorithm}

\subsection{Distributed bucket construction}
\label{sec: Qutanile lookup}

The commonest way for distributed bucket construction in traditional distributed GBDT~\cite{XGBoost, vero, lightGBM, DimBoost} is named \textit{quantile sketch}~\cite{sketch1, sketch2}, which requires each participant to send representations of its local data so that the distribution of each feature can be approximated.
This approach will inevitably reveal information about participants' local data.  
Therefore, we propose a new method for distributed bucket construction so that privacy can be protected.

The basic idea for our distributed bucket construction method is to find the splits (named \textit{quantiles}) that divide $n$ sample values of a feature into $q$ buckets; then, participants put their samples into the corresponding buckets based on these quantiles.
The pseudo-code for finding all $q-1$ quantiles of a feature is shown in Protocol~\ref{alg:quantile}.
We use $\Cli_l$ as the active participant to coordinate the protocol, but any participant can be the active participant. 

For the first quantile, we use binary search to find a value that is larger than $\frac{n}{q}$ sample values and smaller than the rest.
In more detail, $\Cli_l$ initializes two values: $Q_{min}$ and $Q_{max}$, which are the smallest and largest possible values of the feature (line 2).
Then, it initializes $Q_1$ as the mean of $Q_{min}$ and $Q_{max}$ (line 5). 
$\Cli_l$ needs to count the number of sample values ($n'$) that are smaller than $Q_1$. 
By comparing $n'$ to $\frac{n}{q}$, $\Cli_l$ could decide whether to increase $Q_1$ or decrease it for the next round of binary search (line 10-14).

However, $\Cli_l$ is not able to count $n'$, as the samples are distributed among $l$ participants. 
A naive solution is to have participants count locally, and $\Cli_l$ aggregates the results.
Unfortunately, this will reveal information about a participant's local dataset.
To this end, we have all participants aggregate their counts via secure aggregation (line 9).
After finding $Q_1$, participants locally remove\footnote{They remove the values only for this protocol, but still keep them in their datasets.} their sample values that are smaller than $Q_1$ (line 17). 
Then, they find the second quantile $Q_2$ in exactly the same way as for finding $Q_1$.
After finding all quantiles, each $\Cli_i$ knows how to put these samples into the corresponding buckets for this feature, 
and they can find the quantiles for other features in the same way.
We remark that multiple instances of Protocol~\ref{alg:quantile} could run in parallel so that the quantiles of multiple features could be found at the same time.

The method described above only applies to continuous features. 
For discrete features, the number of classes may be less than the number of buckets, and lines 4-15 could be an endless loop. 
In this case, we simply build a bucket for each class. 
Then, we can directly move to the training phase. 



\subsection{Training}
\label{sec:horizontaltrain}

Similar to vertical $\name$, bucket construction in horizontal $\name$ also only needs to be done once: participants can run the training phase multiple times to fine-tune the model without further bucket construction as long as the data remains unchanged.  

After finding all quantiles, each participant can locally put their sample IDs into the corresponding buckets;
$\Cli_l$ can collect the buckets and aggregate them. 
Then, the setting becomes similar to vertical $\name$.
However, $\Cli_l$ does \textit{not} hold all labels; hence it cannot train the decision trees as vertical $\name$.
To this end, we take another approach,
the pseudo-code of which is shown in Protocol~\ref{alg:horizontal}.
The key difference is that instead of sending the buckets of sample IDs to $\Cli_l$,
each $\Cli_i$ locally computes $g$s and $h$s for each sample (line 9), computes $G$s and $H$s for each bucket (line 18) and all participants aggregates $G$s and $H$s for the corresponding buckets using secure aggregation (line 20).
We remark that all $m\cdot q$ instances of secure aggregation can run in parallel.
Another difference is that $\Cli_l$ needs to send the split information to \textit{all} the other participants (line 24).

Naively, we can use the secure aggregation (cf. Section~\ref{sec:aggregation}) protocol~\cite{bonawitz2017practical} by having $\Cli_l$ play the role of the parameter server.
However, we 
simplify the protocol based on the assumption that the participants will \textit{not} drop out in our setting.
The pseudo-code is shown in Protocol~\ref{alg:aggregation}.
Notice that the gradients $G$s are floating-point numbers. 
To deal with this, we scale the floating-point numbers up to integers by multiplying the same constant by all values and dropping the fractional parts. 
This idea is widely used in neural network training and inferences~\cite{secureML, minionn}.
$N$ must be large enough so that the absolute value of the final sum $G$ is smaller than $\left\lfloor N/2 \right\rfloor$. 
\textlabel{We}{Del3} separately detail the security analysis of horizontal $\name$ from the perspective of $\Cli_i$ and $\Cli_l$ in Section~\ref{sec:horizontal_analysis}.

\revision{\Paragraph{Prediction.}
\textlabel{The}{RA:1:3} prediction phase for horizontal $\name$ can be achieved locally by each participant, as all participants hold the trained model in the end.
Recall that $\Cli_l$ sends the split information to all participants (line 24 of Protocol~\ref{alg:horizontal}), based on which the participants know how the buckets are split for all nodes. 
The threshold values are the quantiles that are known to all participants as well. Hence each participant knows the threshold values for nodes of all trees.
}

\subsection{Security analysis}
\label{sec:horizontal_analysis}
\PParagraph{$\Cli_i$'s security.}
There are two places for potential information leakage:
\begin{packeditemize}
    \item During quantile lookup, $\Cli_i$ inputs $n'_i$ to secure aggregation (line 9 of Protocol~\ref{alg:quantile}). 
    \item During tree construction, $\Cli_i$ inputs $G^i_{j,k}$ and $H^i_{j,k}$ to secure aggregation (line 20 of Protocol~\ref{alg:horizontal}).
\end{packeditemize}

Both inputs are protected by secure aggregation.
Although $\Cli_l$ can collude with $\tau-1$ passive participants, it still cannot learn anything beyond the sum of $l-\tau$ participants' inputs.  


\Paragraph{$\Cli_l$'s security.}
There are again two places for potential information leakage:
\begin{packeditemize}
    \item During quantile lookup, $\Cli_l$ sends $Q_j$ to other $\Cli$s (line 5 of Protocol~\ref{alg:quantile}). 
    \item During tree construction, $\Cli_l$ sends $\mathit{split}_{j}$ to other $\Cli$s (line 24 of Protocol~\ref{alg:horizontal}).
\end{packeditemize}
Notice that $Q_j$ is calculated based on $n'$ and $\mathit{split}_j$ is calculated based on $G^i_j$ and $H^i_j$.
Therefore, the information leakage of $\Cli_l$ will not be larger than $\Cli_i$ ($\Cli_i$'s security was proved above).

\section{Implementation and Experiments}
\label{sec:experiments}

In this section, we evaluate $\name$ by conducting experiments on three public datasets:

\begin{packeditemize}
    \item $\textbf{Credit 1}$\footnote{\url{https://www.kaggle.com/c/GiveMeSomeCredit/overview}}: 
    It is a credit-scoring dataset used to predict the probability that somebody will experience financial distress in the next two years.
    It consists of a total of \numprint{150000} samples and \numprint{10} features, \revision{\textlabel{and}{RA:2:2:1} about $6.68\%$ samples are positive}.
    \item $\textbf{\textlabel{Credit} 2}$\footnote{\revision{\url{https://www.kaggle.com/uciml/default-of-credit-card-clients-dataset}}}: 
    It is another credit-scoring dataset correlated to the task of predicting whether a user will make a payment on time. 
    It consists of a total of \numprint{30000} samples and \numprint{23} features, \revision{\textlabel{and}{RA:2:2:2} $22.12\%$ samples are positive}.
    \item $\textbf{SUSY}$\footnote{\url{https://www.csie.ntu.edu.tw/~cjlin/libsvm/}}: It is a dataset about high-energy physics, used to distinguish between a process where new super-symmetric particles are produced leading to a final state in which some particles are detectable, and others are invisible to the experimental apparatus. The original dataset consists of \numprint{3000000} samples, and we choose \numprint{290000} samples randomly from the dataset.  Each sample has \numprint{18} features, \revision{\textlabel{and}{RA:2:2:3} about $45.7\%$ samples are positive.}

\end{packeditemize}

For each dataset, we divide it into two parts for training and testing, respectively.
The training part contains two-thirds of the samples, and the testing part has the remaining one-third. We use the commonly used \revision{\textbf{\textlabel{Area}{RA:2:2:4} under the ROC curve (AUC)}} as the evaluation metric since the negative samples accounted for most of the samples in the Credit 1 dataset.

Our evaluation consists of two parts: utility and efficiency. 
Recall that all participants jointly run the GBDT training algorithm in both vertical and horizontal $\name$,
hence varying the number of participants will not affect the utility of $\name$.
When evaluating utility, we only consider different numbers of buckets and different levels of DP.
We consider different numbers of participants when evaluating efficiency.
For ``Credit 1'' and ``Credit 2'', we set the number of trees as $T=20$ and each tree has 3 layers; 
for ``SUSY'', we set the number of trees as $T=60$ and each tree has 4 layers.
All experiments were repeated 5 times and the averages are reported.

\subsection{Utility}
\label{sec:utility}

We first evaluate the utility of $\name$ with different numbers of buckets.
Then, we fix the number of buckets with an optimal value and run vertical $\name$ with different levels of DP (recall that DP is not needed for horizontal $\name$).

\begin{figure*}[htbp]
    \centering
    \subfigure[Vertical $\name$.]{
        \begin{minipage}[t]{0.29\linewidth}
        \includegraphics[width= \linewidth]{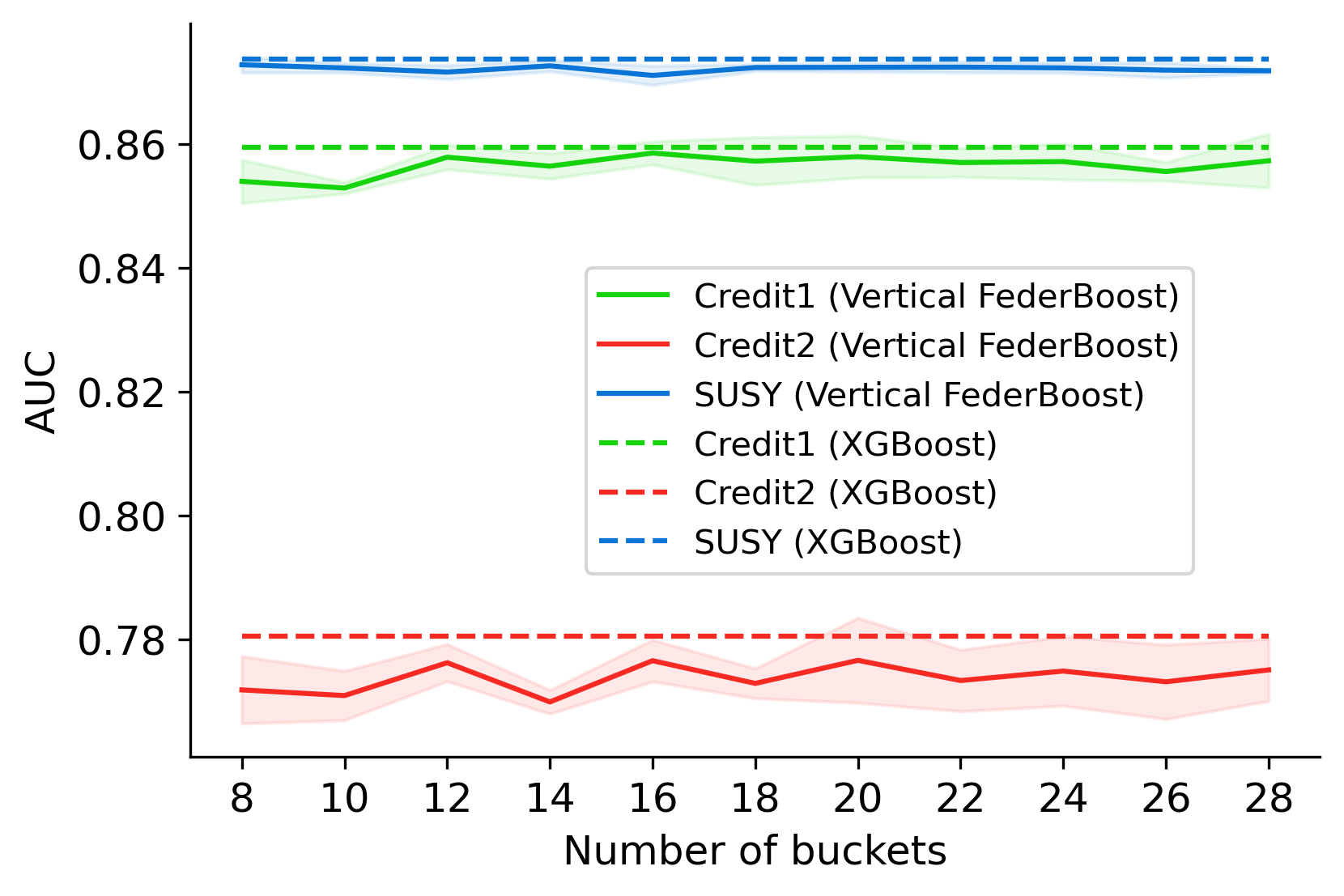}
        \label{fig:plot1}
        \end{minipage}
    }
    \subfigure[Horizontal $\name$.]{
        \begin{minipage}[t]{0.29\linewidth}
        \includegraphics[width= \linewidth]{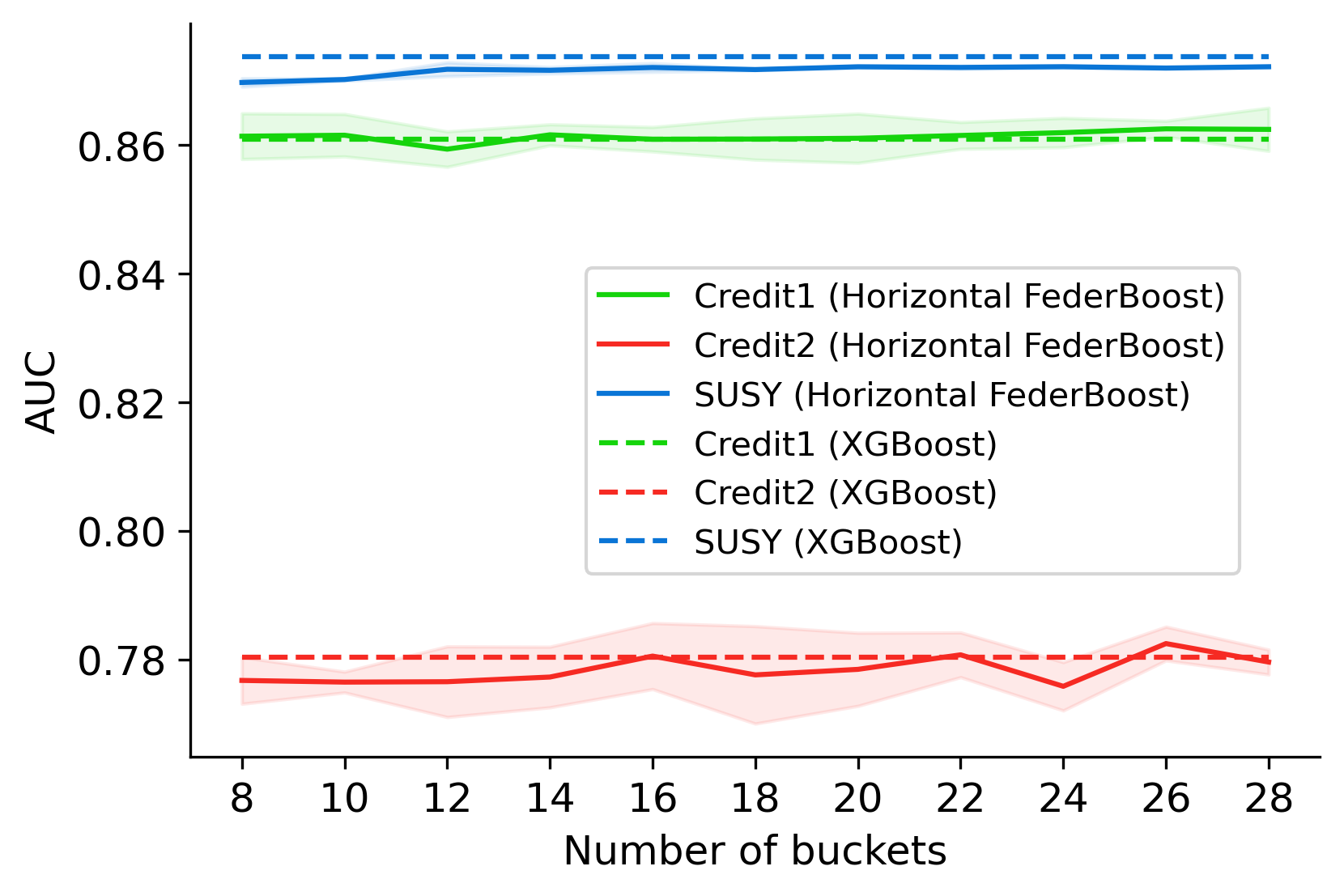}
        \label{fig:plot2}
        \end{minipage}
    }
    \rulesep
    \subfigure[Vertical $\name$ under DP.]{
        \begin{minipage}[t]{0.29\linewidth}
        \includegraphics[width= \linewidth]{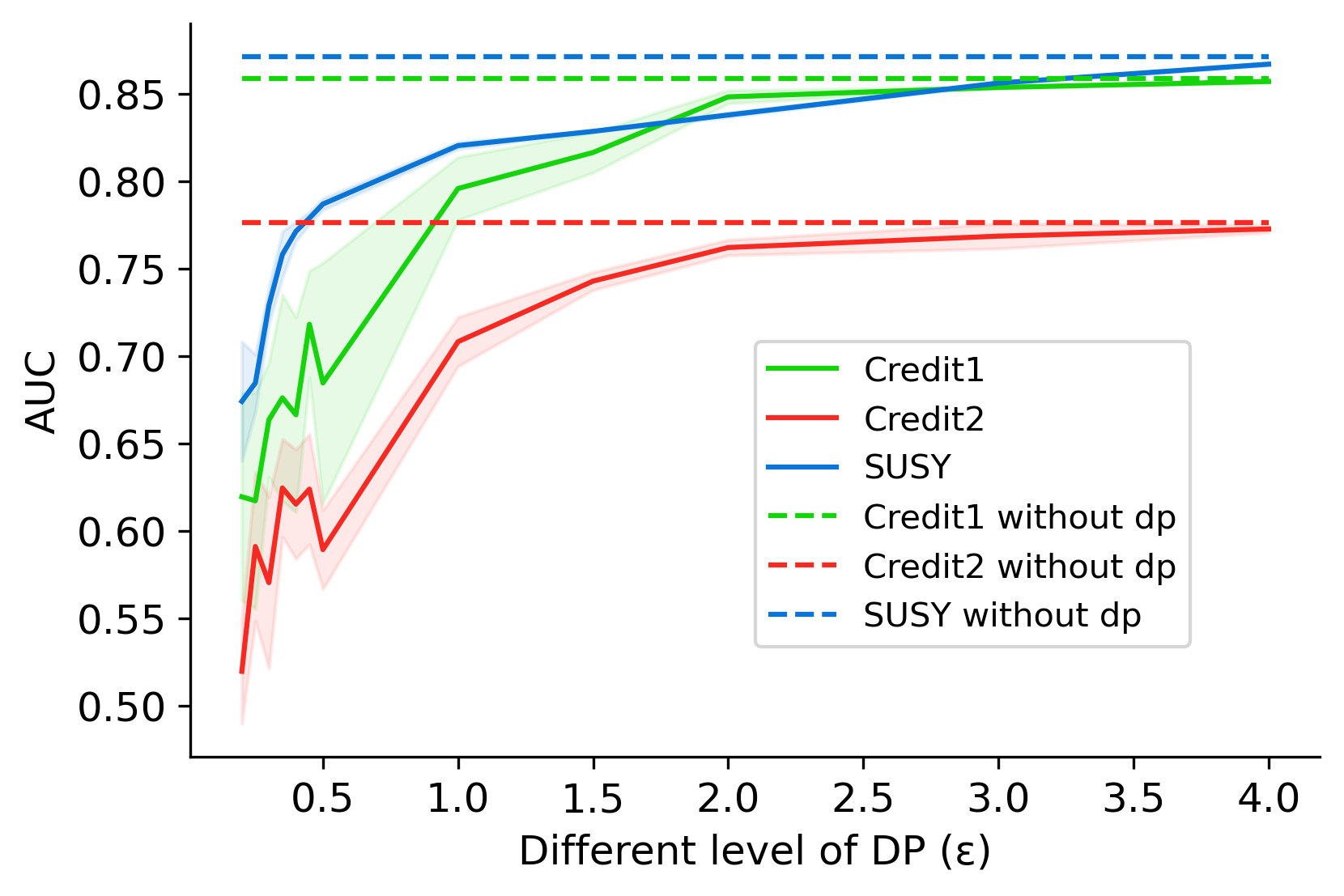}
        \label{fig:DP}
        \end{minipage}
    }
    \caption{Utility of $\name$ with different number of buckets (Left) and different levels of DP (Right)}
    \label{fig:buckets}
\end{figure*}

We also run XGBoost centrally with the same datasets and use the results as baselines.
As shown in \fig~\ref{fig:buckets}, {\bf $\name$ achieves almost the same accuracy with XGBoost}. 
For ``Credit 1'', the AUC achieved by XGBoost is {\bf 86.10}\% ;
the best AUC achieved by vertical $\name$ is {\bf 85.85}\% with 16 buckets;
and the best AUC achieved by horizontal $\name$ is {\bf 86.25}\% with 26 buckets. 
For ``Credit 2'', the AUC achieved by XGBoost is {\bf 78.04}\% ;
the best AUC achieved by vertical $\name$ is {\bf 77.65}\% with 16 buckets;
and the AUC achieved by horizontal $\name$ is {\bf 78.25}\% with 24 buckets. 
For ``SUSY'', the AUC achieved by XGBoost is {\bf 87.37}\%;
the AUC achieved by vertical $\name$ is {\bf 87.26}\% with 14 buckets;
and the AUC achieved by horizontal $\name$ is {\bf 87.22}\% with 24 buckets. 

The performance of horizontal $\name$ is better than vertical $\name$ since we bucketize samples according to their quantile, which can better characterize the distribution of data. In contrast, samples are equally partitioned into different buckets in vertical $\name$. 
Unfortunately, we cannot adopt the same strategy in vertical $\name$ since this will leak more information: in vertical $\name$, each passive participant sends the sample IDs in each bucket to the active participant. In contrast, horizontal $\name$ only requires inputting the gradient of each bucket to secure aggregation. 



Next, we set the number of buckets as 16 and run vertical $\name$ with different levels of DP. 
The results in \fig~\ref{fig:DP} shows that
\textbf{when $\epsilon=4$, the accuracy achieved by vertical $\name$ is very close to that without DP for all three datasets}.
For ``Credit 1'', vertical $\name$ achieves {\bf 85.70}\% accuracy when $\epsilon=4$ (85.85\% when no DP added).
For ``Credit 2'', vertical $\name$ achieves {\bf 77.27}\% accuracy when $\epsilon=4$ (77.65\% when no DP added).
For ``SUSY'', vertical $\name$ achieves {\bf 86.69}\% accuracy when $\epsilon=4$ (87.10\% when no DP added).



\subsection{Efficiency}
\label{sec:efficiency}

We fully implement $\name$ in C++ using GMP\footnote{https://gmplib.org/} for cryptographic operations.
We deploy our implementation on a machine that contains 40 2.20GHz CPUs and 251 GB memory; we spawn up to 32 processes, and each process runs as a single participant.
\revision{\textlabel{We}{RA3:2} utilize the traffic control~\cite{tc} integrated within the Linux kernel to set the traffic conditions among processes to simulate the authentic network communication among participants. By configuring the bandwidth, latency, and other parameters, we strive to recreate the real network environment between participants. Such an approach has been extensively utilized in previous research works~\cite{DBLP:journals/iacr/AbspoelEV20, DBLP:journals/pvldb/WuCXCO20,DBLP:conf/uss/HuangLHD22} and experiments show that it can effectively reflect the communication overhead comparison of different protocols in real environments.
}
\deleted{For communication overhead} 
\revision{Specifically}, we consider both the local area network (LAN) and the wide area network (WAN) \revision{in our experiments}. 
\deleted{To  simulate  WAN, we  limit  the  network  bandwidth  of  each  process  to  20 Mbit/s and  add  100ms  latency  to  each  link  connection.}
\revision{We constrain each process to a network bandwidth of 1000 Mbit/s and introduce a 0.1 ms latency to each link connection to simulate LAN. To simulate WAN, we limit  the  network  bandwidth  of each  process  to  20 Mbit/s and  add  100 ms  latency  to  each  link  connection. In order to facilitate comparison with other methodologies, we adjust our network environment settings to match those reported in those methods.}

\begin{figure*}[!htb]
    \centering
    \subfigure[Vertical $\name$]{
        \begin{minipage}[t]{0.23\linewidth}
        \includegraphics[width=\linewidth]{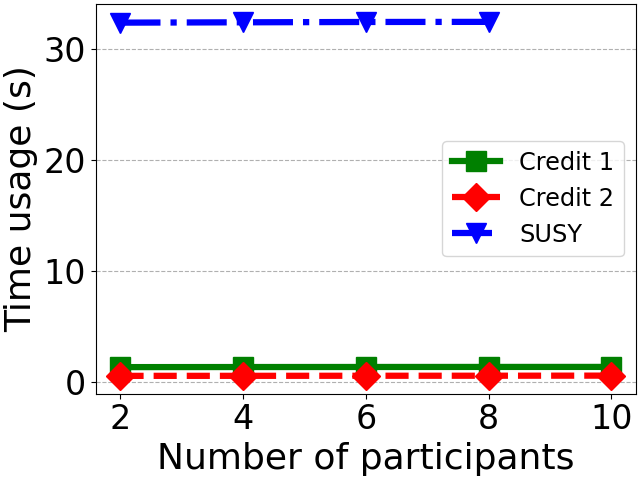}
        \label{fig:train_time_vertical}
        \end{minipage}
    }
    \subfigure[Horizontal $\name$]{
        \begin{minipage}[t]{0.23\linewidth}
        \includegraphics[width=\linewidth]{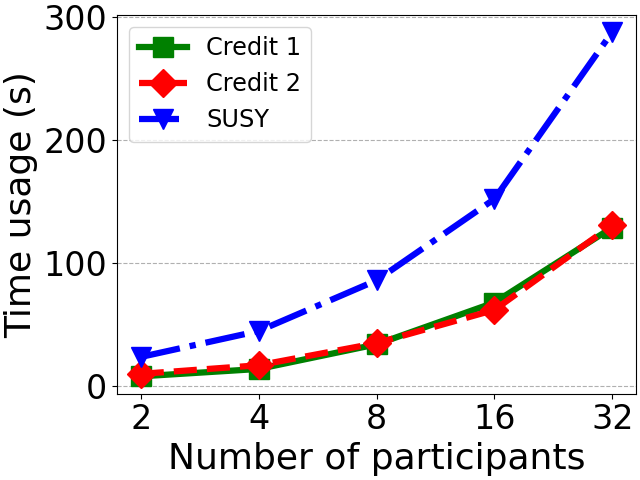}
        \label{fig:train_time_horizontal}
        \end{minipage}
    }
    \rulesep
    \subfigure[Vertical $\name$]{
        \begin{minipage}[t]{0.23\linewidth}
        \includegraphics[width=\linewidth]{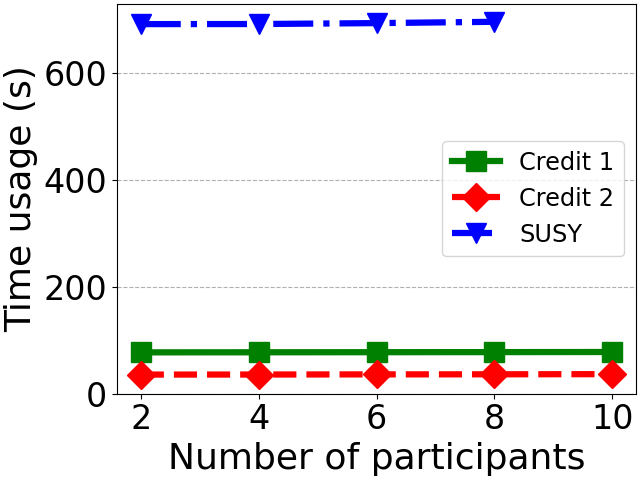}
        \label{fig:train_time_vertical_delay}
        \end{minipage}
    }
    \subfigure[Horizontal $\name$]{
        \begin{minipage}[t]{0.23\linewidth}
        \includegraphics[width=\linewidth]{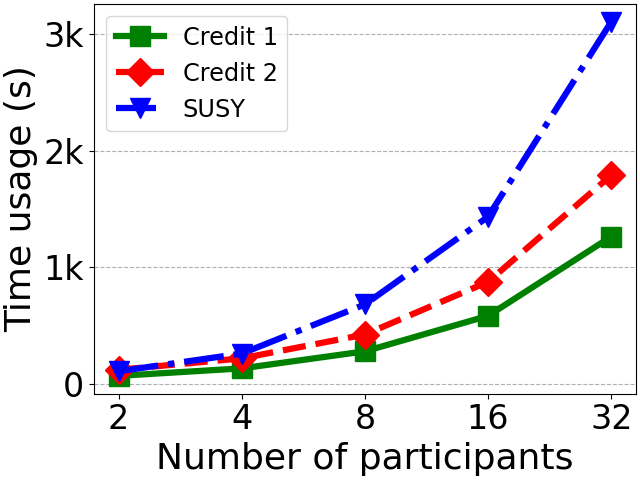}
        \label{fig:train_time_horizontal_delay}
        \end{minipage}
    }
    \caption{Training time of $\name$ with different number of participants w.r.t. LAN (Left) and WAN (Right)}
    \label{fig:participants}
\end{figure*}

\fig~\ref{fig:train_time_vertical} shows the training time of vertical $\name$ in LAN with different number of participants. 
The results show that vertical $\name$ is very efficient: even for the challenging ``SUSY'' dataset, it only takes at most \textbf{33 seconds} to train a GBDT model.

\fig~\ref{fig:train_time_horizontal} shows the training time of horizontal $\name$ in LAN. 
Both the bucket construction phase and the training phase
require secure aggregation for each quantile lookup and each tree node split, respectively.
Recall that the efficiency of secure aggregation depends on the number of participants. Hence the time usage of horizontal $\name$ increases linearly with the number of participants. 
For the ``SUSY'' dataset, it takes {\bf 23.75 seconds} for 2 participants and {\bf 86 seconds} for 8 participants.
The results for ``Credit1'' and ``Credit2'' are similar: around {\bf 10 seconds} for 2 participants and {\bf 130 seconds} for 32 participants.

\fig~\ref{fig:train_time_vertical_delay} shows the training time of vertical $\name$ in WAN. 
In ``SUSY'', it takes at most {\bf 691.77 seconds} to train a GBDT model. 
Compared to LAN, the training time increases significantly because passive participants need to transfer buckets of IDs to the active participant, which is expensive in WAN. 
\fig~\ref{fig:train_time_horizontal_delay} shows the training time of horizontal $\name$ in WAN.
It takes at most {\bf {3103.24} seconds} to finish training. 

Notice that the bucket construction phase only needs to be done once in either vertical $\name$ or horizontal $\name$.
It occupies more than half of the total time usage.
If we remove bucket construction from the results, it will show a significant speedup. 

We also compare $\name$ with Abspoel et al.~\cite{DBLP:journals/iacr/AbspoelEV20} and Wu et al.~\cite{DBLP:journals/pvldb/WuCXCO20}, which are the state-of-the-art solutions for federated decision tree training. 
Abspoel et al.~\cite{DBLP:journals/iacr/AbspoelEV20} is based on three party honest-majority replicated secret sharing;
in particular, they use oblivious sorting to sort the samples for each feature. 
This scheme supports both vertically and horizontally partitioned data, but can only support three participants. They simulated each participant using a 2.5 GHz CPU. Moreover, their benchmarks were conducted in LAN with a dataset consisting of \numprint{8192} samples and 11 features;
they train 200 trees and each tree has 4 layers. We tailor our SUSY dataset to the same dimension and evaluate vertical and horizontal $\name$ in the same setting. 
Table~\ref{tab:comparsion} shows our time usage compared with the results reported in Table 1 of~\cite{DBLP:journals/iacr/AbspoelEV20}.
{\bf Vertical $\name$ achieves \numprint{83099} times speedup and horizontal $\name$ achieves \numprint{4668} times speedup.}

\begin{table}[htb]
\caption{FederBoost vs. Abspoel et al.
\label{tab:comparsion}
\cite{DBLP:journals/iacr/AbspoelEV20}\\
(200 4-layer trees, 8192 samples, 11 features, 3 participants)}
\begin{tabularx}{8.5cm}{@{} lX} 
\toprule 
\textbf{Method} &  \textbf{Time usage}
\\
\midrule  
\ Vertical $\name$ &  1.213 s \\ 
\ Horizontal $\name$ &  21.59 s\\ 
\ Abspoel et al.~\cite{DBLP:journals/iacr/AbspoelEV20}  & $\sim$28 hours \\
\bottomrule 
\addlinespace
\end{tabularx}
\end{table}  

Wu et al.~\cite{DBLP:journals/pvldb/WuCXCO20} combine threshold partially homomorphic encryption (TPHE) with MPC.
This scheme only supports vertically partitioned data. 
They simulated each participant using a 3.5 GHz CPU. Their benchmarks were conducted in LAN with a dataset consisting of \numprint{50000} samples and 15 features;
they train up to 32 trees and each tree has 4 layers. 
\textlabel{We}{Del4} also tailor our SUSY dataset to the same dimension and evaluate vertical $\name$ in the same setting. 
Table~\ref{tab:comparsion2} shows our time usage compared with the results reported in Figure 4(f) of~\cite{DBLP:journals/pvldb/WuCXCO20}.
\textbf{$\name$ achieves \numprint{1111} times speedup.}

\begin{table}[htb]
\caption{FederBoost vs. Wu et al.
\cite{DBLP:journals/pvldb/WuCXCO20}\\
(32 4-layer trees, \numprint{50000} samples, 15 features, 3 participants)}
\label{tab:comparsion2}
\begin{tabularx}{8.5cm}{lX}  
\toprule 
\textbf{Method} &  \textbf{Time usage}
\\
\midrule  
\ Vertical $\name$ &  32.4 s \\ 
\ Wu et al. \cite{DBLP:journals/pvldb/WuCXCO20}  & $\sim$10 hours \\
\bottomrule 
\end{tabularx}   
\end{table}




\section{Related Work}
\label{sec:related_work}
In addition to the solutions proposed by Abspoel et al.~\mbox{\cite{DBLP:journals/iacr/AbspoelEV20}} and Wu et al.~\mbox{\cite{DBLP:journals/pvldb/WuCXCO20}} (cf. Section~{\ref{sec:efficiency}}), there are some other works that solve the problem of federated decision tree training.
Even though not specifically mentioned, the first federated decision tree learning algorithm was proposed by Lindell and Pinkas in 2000~\mbox{\cite{Benny}}. 
They came up with a protocol allowing two participants to privately compute the ID3 algorithm over horizontally partitioned data. 
Recently, Cheng et al.~\mbox{\cite{secureboost}} propose SecureBoost, a federated GBDT framework for vertically partitioned data.
This protocol requires cryptographic computation and communication for each possible split, hence is expensive. 
As a comparison, our vertical $\name$ does not require any cryptographic operation. Chen et al.~\mbox{\cite{DBLP:journals/corr/abs-2110-10927}} incorporates several engineering optimizations into SecureBoost. Experiments on the Credit2 dataset show that it requires at least 30 seconds to train a single tree, while we only require 2 seconds to train 20 trees.

Another recent work~\mbox{\cite{Wenting}} for federated GBDT was achieved using trusted execution environments (TEEs)~\cite{AMD_trustzone, SGX}.
It introduces a central server that is equipped  with a TEE.
All participants send their data, no matter whether vertically or horizontally partitioned, to the TEE via secure channels.
However, TEEs are known to be vulnerable to hardware-based side-channel attacks~\cite{Foreshadow}.
Alternatively, Li et al.~\mbox{\cite{DBLP:journals/corr/abs-1911-04206}} apply locality-sensitive hashing (LSH) to federated GBDT. However, their solution only supports horizontally partitioned data, and the security of LSH is difficult to quantify. 
Zhu et al.~\mbox{\cite{DBLP:journals/corr/abs-2108-11444}} considers a setting where the data is vertically partitioned, but the labels are distributed among multiple clients, whereas we assume the labels are stored only on one client. Furthermore, they only protect the privacy of labels, whereas we protect both data and labels.

\section{\revision{Discussion}}
\label{ssec:discuss}
\PParagraph{\revision{Generalization of $\name$.}}
\revision{The direct application of $\name$ to other machine learning-based methods may pose significant challenges. Nevertheless, certain aspects of our approach have the potential to be extended to other methods. Notably, our utilization of secure aggregation to aggregate gradients from passive participants in horizontal $\name$  can serve as a helpful tool for federated training of neural network models, where each client sends their local gradients to a centralized server to get the global gradients. Such an idea has already been applied in previous research efforts~\cite{bonawitz2016practical, mcmahan2023communicationefficient}. Additionally, our method of vertical $\name$  can be applied to other approaches that rely on knowledge of data order during training. For example, transformer-based language models~\cite{vaswani2017attention} that encode the position of each token may benefit from our approach.}

\section{Conclusion}

In response to the growing demand for a federated GBDT framework,
we propose $\name$ that supports running GBDT privately over vertically and horizontally partitioned data. 
Vertical $\name$ does not require any cryptographic operation, and horizontal $\name$ only requires lightweight secure aggregation.
Our experimental results show that both vertical and horizontal $\name$ achieves the same level of accuracy with  centralized training, and they are 4-5 orders of magnitude faster than the state-of-the-art solution for federated decision tree \textlabel{training}{Del6}.


In future work, we will further improve the performance of $\name$ by improving communication among participants through structured networks~\mbox{\cite{DBLP:journals/iacr/BellBGLR20}} and addressing potential poisoning attacks. Additionally, we will attempt to deploy $\name$ in real industrial scenarios to test its effectiveness on realistic data.


\section*{Acknowledgments}

The work was supported in part by 
National Natural Science Foundation of China (Grant No. 62002319, U20A20222),
Hangzhou Leading Innovation and Entrepreneurship Team (TD2020003),
and
Zhejiang Key R\&D Plan (Grant No. 2021C01116).



\bibliographystyle{IEEEtran}
\bibliography{main}

\begin{IEEEbiography}[{\includegraphics[width=1in,height=1.25in,clip,keepaspectratio]{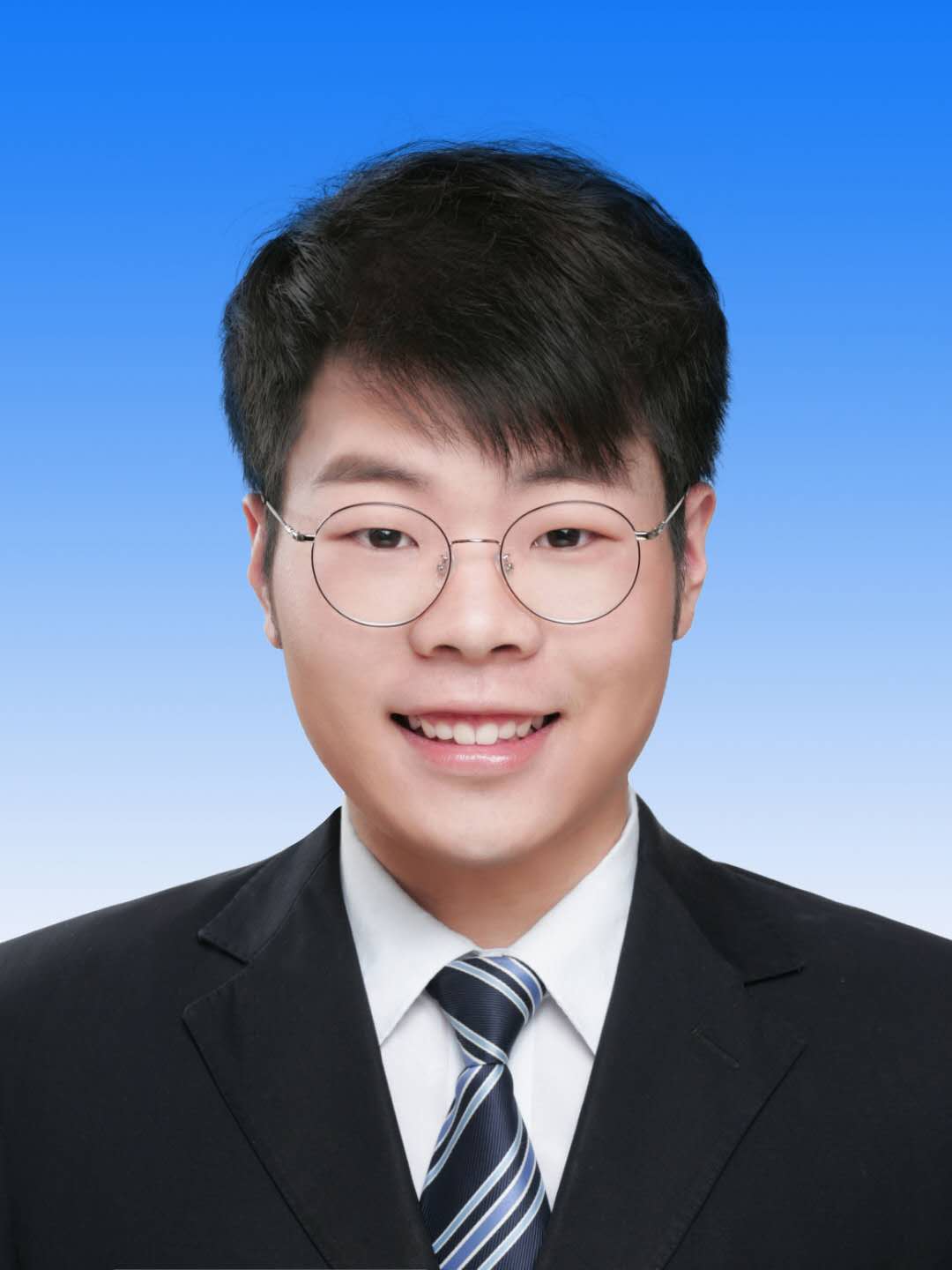}}]{Zhihua Tian} received the B.Sc. degree in stastics from Shandong University, Ji’nan, China, in 2020. Currently, he is pursuing the Ph.D. degree in the School of Cyber Science and Technology of Zhejiang University, Hangzhou, China. His
research interests include machine learning, federated learning and adversarial training algorithms. 
\end{IEEEbiography}

\begin{IEEEbiography}[{\includegraphics[width=1in,height=1.25in,clip,keepaspectratio]{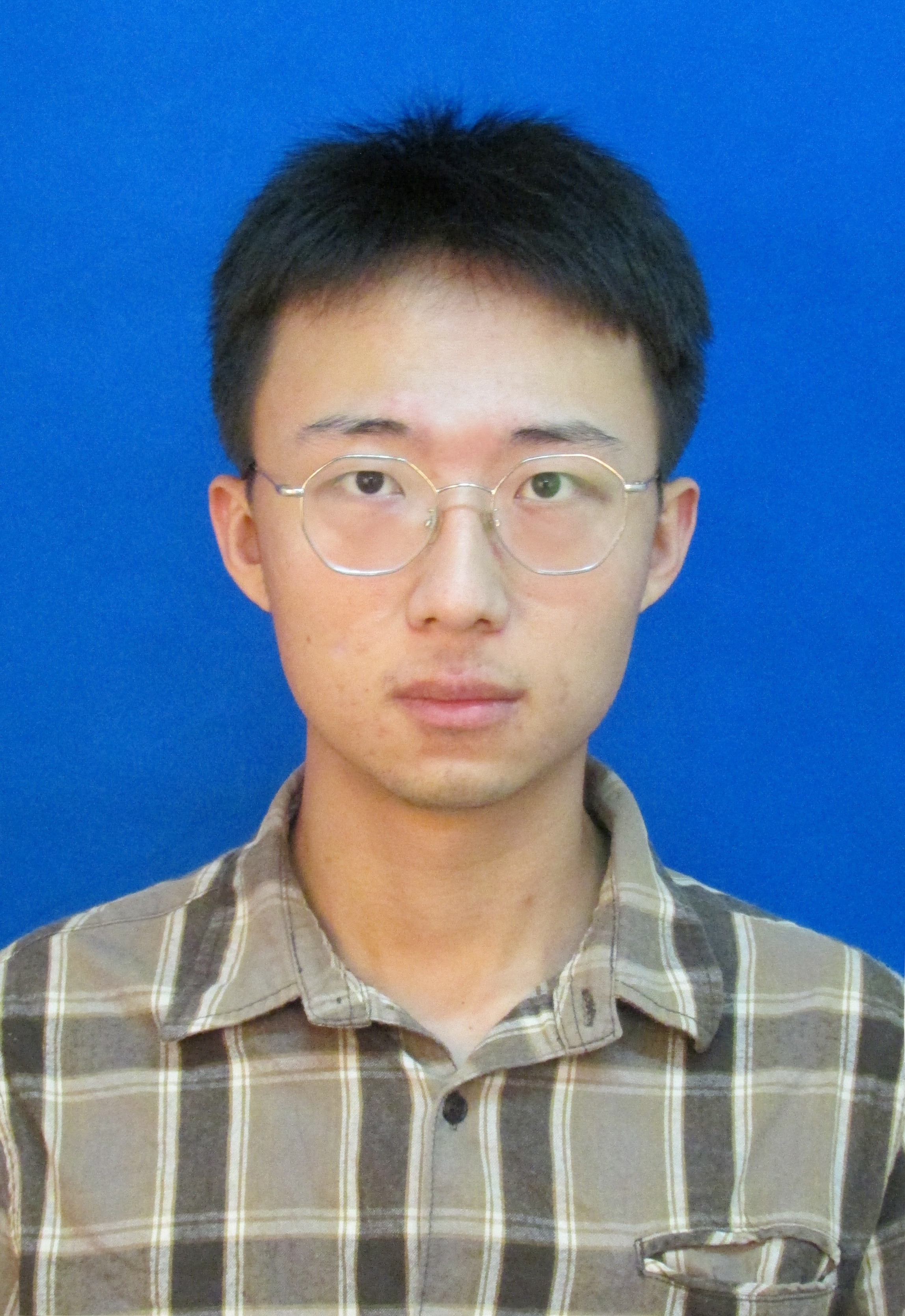}}]{Zhang Rui} graduated from School of Mathematics in Zhejiang University, Hangzhou China, in 2020. Currently, he is working on his PhD in School of Mathematics of Zhejiang University, Hangzhou China. His research interests include machine learning, federated learning and adversarial training algorithms.
\end{IEEEbiography}

\begin{IEEEbiography}[{\includegraphics[width=1in,height=1.25in,clip,keepaspectratio]{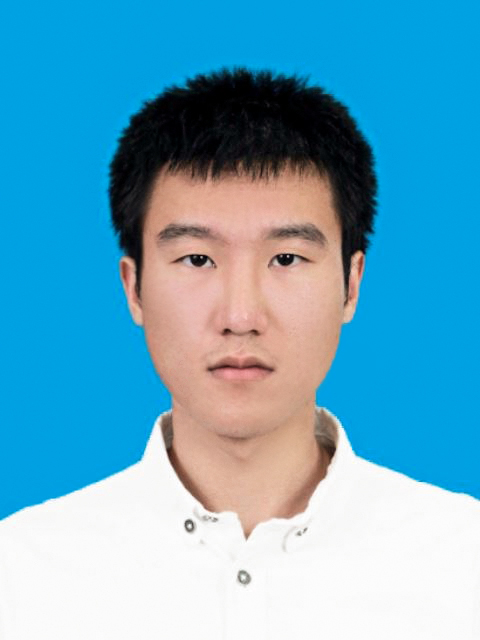}}]{Xiaoyang Hou} is a Ph.D. student at Zhejiang University. Before that, he received his bachelor's degree in 2021 from Zhejiang University of Technology. His research is on federated learning and applied cryptography.
\end{IEEEbiography}

\begin{IEEEbiography}[{\includegraphics[width=1in,height=1.25in,clip,keepaspectratio]{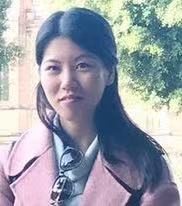}}]{Lingjuan Lyu} is a senior research scientist and Privacy-Preserving Machine Learning (PPML) team leader in Sony AI. She received Ph.D. from the University of Melbourne. She was a winner of the IBM Ph.D. Fellowship Worldwide. Her current research interest is trustworthy AI. She had published over 50 papers in top conferences and journals, including NeurIPS, ICML, ICLR, Nature, AAAI, IJCAI, etc. Her papers had won several best paper awards and oral presentations from top conferences.
\end{IEEEbiography}

\begin{IEEEbiography}[{\includegraphics[width=1in,height=1.25in,clip,keepaspectratio]{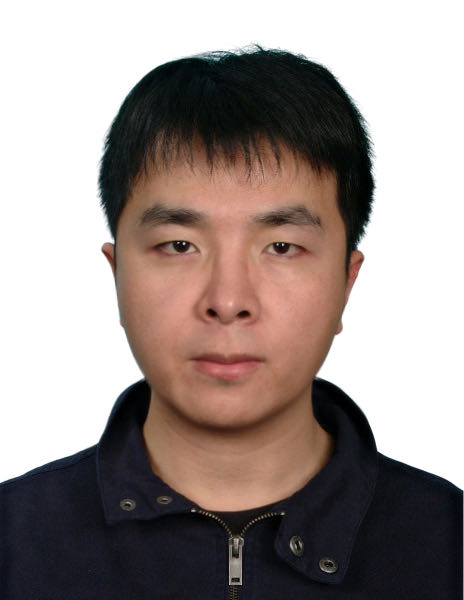}}]{Tianyi Zhang} received his Master's degree in Computer Science from Boston University, USA, in 2021. He is currently employed as a Software Engineer at Amazon, where he applies his technical expertise and problem-solving skills to develop innovative software solutions. His current research interests include artificial intelligence, machine learning, and cloud computing technologies.
\end{IEEEbiography}

\begin{IEEEbiography}[{\includegraphics[width=1in,height=1.25in,clip,keepaspectratio]{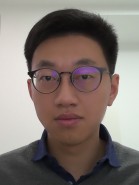}}]{Jian Liu}
is a ZJU100 Young Professor at Zhejiang University. Before that, he was a postdoctoral researcher at UC Berkeley. He got his PhD in July 2018 from Aalto University. His research is on Applied Cryptography, Distributed Systems, Blockchains and Machine Learning. He is interested in building real-world systems that are  provably secure, easy to use and inexpensive to deploy.
\end{IEEEbiography}

\begin{IEEEbiography}[{\includegraphics[width=1in,height=1.25in,clip,keepaspectratio]{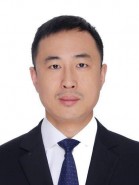}}]{Kui Ren}
is the director of the Institute of Cyberspace Research of Zhejiang University, ACM/IEEE Fellow.
He chaired and participated in projects from NSFC, NSF, US Department of Energy and Air Force Research Laboratory which cost more than \$10 million.
He was awarded the Achievement Award of IEEE ComSoc, Sustained Achievement Award of exceptional scholar program of Suny Buffalo, and  NSF youth Achievement Award. 
Prof. Kui Ren is internationally recognized for accomplishments in the areas of cloud security and wireless security. 
He has made many fundamental innovations in both theory and practice from the discipline to the society. 
His works have significant impact on emerging cloud systems and wireless network technologies.
Prof. Kui Ren has published more than 200 papers, and cited more than 20 thousand times. 
His research results have been widely reported by Xinhua news agency, Scientific American, NSF news, ACM news and other media.

\end{IEEEbiography}
\vfill

\end{document}